\documentclass[%
,aps%
 ,twocolumn%
 ,secnumarabic%
 ,superscriptaddress%
,amssymb, amsmath,nobibnotes, aps, prl, floatfix]{revtex4-2}
\usepackage{docs}%
\usepackage{bm}%
\RequirePackage[utf8]{inputenc}
\RequirePackage{xspace}
\RequirePackage{graphicx}
\RequirePackage{amsmath}
\RequirePackage{amssymb}
 \RequirePackage{xcolor}
 \RequirePackage[sort&compress]{natbib}
 \RequirePackage{hepunits}
 \RequirePackage{enumitem}
\RequirePackage{esint}
 \usepackage{tikz}
 \usepackage[compat=1.1.0]{tikz-feynman}
\usepackage[colorlinks=true,linkcolor=blue]{hyperref}%
 \usepackage{definitions}%
 \usepackage{comments}%
 \usepackage{orcidlink}
 \usepackage{CJKutf8}
 \usepackage{changes}

\begin{document}

\title{Constraining the Energy Momentum Tensor through DVCS Dispersion Relation beyond Leading Power}
\author{V\'ictor \surname{Mart\'inez-Fern\'andez}\orcidlink{0000-0002-0581-7154}}
\email{victor.martinezfernandez@cea.fr}
\affiliation{Irfu, CEA, Université Paris-Saclay, F-91191, Gif-sur-Yvette, France}
\affiliation{Center for Frontiers in Nuclear Science, Stony Brook University, Stony Brook, NY 11794, USA}
\author{Daniele \surname{Binosi}\orcidlink{0000-0003-1742-4689}}
\email{binosi@ectstar.eu}
\affiliation{European Centre for Theoretical Studies in Nuclear Physics and Related Areas, Villa Tambosi, Strada
delle Tabarelle 286, I-38123 Villazzano (TN), Italy}
\author{Cédric \surname{Mezrag}\orcidlink{0000-0001-8678-4085}}
\email{cedric.mezrag@cea.fr}
\affiliation{Irfu, CEA, Université Paris-Saclay,  F-91191, Gif-sur-Yvette, France}
\author{Zhao-Qian \surname{Yao}\orcidlink{0000-0002-9621-6994} {(\begin{CJK}{UTF8}{gbsn}姚照千\end{CJK})}
}
\email{zhaoqian.yao@dci.uhu.es}
\affiliation{Dpto. Ciencias Integradas, Centro de Estudios Avanzados en Fis., Mat. y Comp., Fac. Ciencias Experimentales, Universidad de Huelva, E-21071 Huelva, Spain}
\affiliation{Dpto. Sistemas F\'isicos, Qu\'imicos y Naturales, Univ. Pablo de Olavide, E-41013 Sevilla, Spain}
\begin{abstract}
  In this letter, we analyse and interpret the kinematic power corrections to deeply virtual Compton scattering dispersion relation.
  We show that the kinematic corrections at twist-4 can be connected to other form factors of the Energy-Momentum Tensor beyond the pressure distribution involved at leading-power, namely the ones related to Momentum and total Angular Momentum distributions.
  In the nucleon case, these corrections are not negligible at presently accessible virtualities.
  The DVCS subtraction constant becomes an experimental constraint on momentum distributions, pressure forces distributions, and total angular momentum distributions.
  Finally, we use continuum and lattice-QCD results to predict the expected size of the DVCS subtraction constant, and conclude that momentum distributions are responsible of roughly one-third of the experimental signal at $Q^2 = 2\textrm{GeV}^2$.
\end{abstract}

\maketitle

\section{Introduction}

In the last decade, the question of spatial mapping of pressure and shear forces within the nucleon \cite{Polyakov:2002yz} has been put at the forefront of hadron physics.
Focusing on the valence quark sector, a large activity both on the theoretical (see, for instance, Ref. \cite{Lorce:2025oot,Ji:2025qax}) and experimental sides \cite{Burkert:2018bqq,Kumericki:2019ddg,Dutrieux:2021nlz} has been dedicated to the nucleon energy-momentum tensor (EMT) and its form factors.
The main argument of these studies is that the EMT form factor $C$, containing the information regarding the pressure and shear forces due to quarks within the nucleon can be indirectly probed through Deeply Virtual Compton Scattering (DVCS) \cite{Ji:1996nm,Ji:1996ek}, bypassing the full extraction of Generalised Parton Distribution (GPDs) thanks to dispersion relations (DRs) \cite{Anikin:2007yh,Diehl:2007jb,Dutrieux:2024bgc}.
At leading twist, the subtraction constant can be fully understood in terms of the Polyakov-Weiss $D$-term \cite{Polyakov:1999gs}, whose first moment yields the $C$ form factor.

However, when taking into account kinematic-power corrections \cite{Braun:2014sta,Martinez-Fernandez:2025rcg}, the $D$-term is not the sole component contributing to the DVCS subtraction constant, and at first sight, the extraction of both GPDs $H$ and $E$ is necessary.
What is more, these power corrections are not negligible at current facilities.
Even when considering the future Electron-Ion Collider, the kinematic range is such that these corrections will be sizeable, threatening the interpretability of the DVCS subtraction constant.

In this letter, we will show that this is in practice not the case, and that, even if the power corrections cannot be neglected, most of their contribution comes from the momentum distribution and the angular momentum distribution.
Thus, the DVCS subtraction constant can be seen as an experimental constraint relating momentum, pressure and angular momentum distributions within the nucleon.
This will provide an experimental benchmark for both continuum and lattice QCD computation of the nucleon EMT form factors.  

\section{DVCS dispersion relations}

\begin{figure}[!t]
  \centering
  \includegraphics[width=0.39\textwidth]{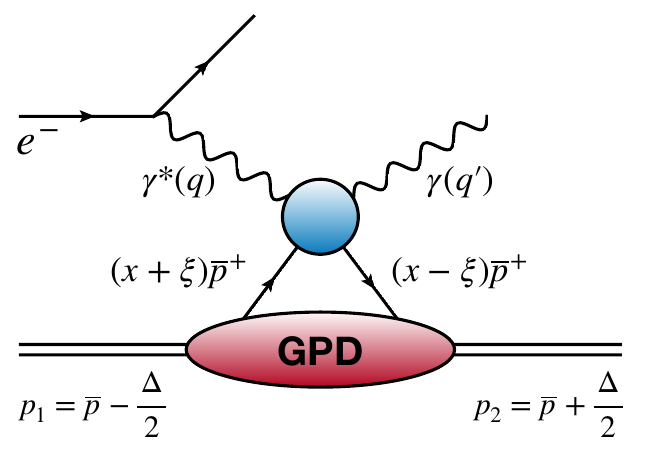}
  \caption{Sketch of the DVCS process factorised into a hard kernel and GPDs. The incoming electron emits a virtual photon of virtuality $Q^2=-q^2$ while a real photon is detected in the final state. The total momentum transfer is $t=\Delta^2$.}
  \label{fig:DVCS}
\end{figure}

The amplitude of the DVCS process, illustrated on Fig.~\ref{fig:DVCS}, can be parameterized in terms of the so-called Compton Form Factors (CFFs) \cite{Belitsky:2001ns,Belitsky:2012ch}. When the photon virtuality $Q^2$ is large enough, the CFFs can be factorised into a hard kernel, that can be computed in perturbation theory, and GPDs containing the non-perturbative information of the process.
The number of GPDs involved in the process description depends on the spin of the target.
Here, we will focus only on the unpolarised, helicity conserving GPDs in the quark sector.
This corresponds to a single GPD $H^q$ for (pseudo-)scalar targets, and a second GPD $E^q$ for spin-$1/2$ targets.

Importantly, GPDs obey a general property called polynomiality \cite{Ji:1998pc,Radyushkin:1998bz}:
\begin{align}
  \label{eq:polynomialityH}
  \int_{-1}^1\ud x\,x^{m} H^q(x,\xi,t) & = \sum_{j=0}^{\left[\frac{m}{2}\right]} A_{m,2j}^q(t) (2\xi)^{2j}\nonumber \\
  &\hspace{-20pt}\quad  +\textrm{mod}(2,m)(2\xi)^{m+1}C_m^q(t),\\
  \label{eq:polynomialityE}
  \int_{-1}^1\ud x\,x^{m} E^q(x,\xi,t) & = \sum_{j=0}^{\left[\frac{m}{2}\right]} B_{m,2j}^q(t) (2\xi)^{2j}\nonumber \\
&\hspace{-20pt}\quad  -\textrm{mod}(2,m)(2\xi)^{m+1}C_m^q(t),
\end{align}
where $[\dots]$ is the floor function and $\textrm{mod}(2,m)$ is 0 for $m$ even and 1 for $m$ odd. $A^q_{1,0}=A^q$, $B^q_{1,0}=B^q$ and $C^q_1=C^q$ are actually the symmetric, current conserving EMT form factors \cite{Ji:1996ek}:
\begin{align}
  \label{eq:EMTDef}
  &\bra{p_2, s'} \texttt{T}^{\{\mu\nu\}}_q(0)\ket{p_1, s}
    = \bar u(p_2, s') \Bigg\{ \frac{\Delta^\mu\Delta^\nu - \eta^{\mu\nu}\Delta^2}{M}\, C^q(t)
    \nonumber
  \\
  & + \frac{\bp^\mu \bp^\nu}{M}\,A^q(t)+\frac{\bp^{\{\mu} i\sigma^{\nu\}\rho}\Delta_\rho}{4M}\left[A^q(t)+B^q(t)\right]\Bigg\} u(p_1, s) \,,
\end{align}
The polynomiality property is equivalent to the existence of the so-called double distribution (DD) formalism introduced independently in \cite{Mueller:1998fv,Radyushkin:1997ki} (see also \cite{Chouika:2017dhe,Mezrag:2022pqk}). 
\begin{align}
  \label{eq:DDF}
  H^q(x,\xi,t) = & \int_\Omega \textrm{d}\Omega \bigg[F^q(\beta,\alpha,t) +\xi D^q(\alpha,t)\delta(\beta) \bigg] \delta_{x;\xi}^{\beta,\alpha},\\
    \label{eq:DDK}
  E^q(x,\xi,t) = & \int_\Omega \textrm{d}\Omega \bigg[ K^q(\beta,\alpha,t)-\xi D^q(\alpha,t)\delta(\beta) \bigg]\delta_{x;\xi}^{\beta,\alpha},
\end{align}
where $\Omega = \{(\alpha,\beta)| |\alpha|+|\beta|\le 1\}$ and $\delta_{x;\xi}^{\beta,\alpha} = \delta(x-\beta-\alpha \xi)$.
$F$ and $K$ are Radyushkin's DD while $D$ is the Polyakov-Weiss term \cite{Polyakov:1999gs}. They are the generating functions of the moments $A_{m,2j}^q(t)$, $B_{m,2j}^q(t)$ and $C_m^q(t)$ respectively.

When the kinematic conditions allow it, the DVCS amplitude obeys subtracted dispersion relations.
The best known one is given as \cite{Anikin:2007yh,Diehl:2007jb,Dutrieux:2024bgc}:
\begin{align}
  \label{eq:DRStandard}
  \mathcal{S} = \mathrm{Re}\, \mathcal{H} - \frac{2}{\pi} \fint_0^1\textrm{d}x  \left(\frac{x}{\xi} \right) \frac{x\, \mathrm{Im}\, \mathcal{H}(x)}{(\xi-x)(\xi+x)},
\end{align}
where $\fint$ represents the Cauchy principal value prescription, $\mathcal{S}$ is the subtraction constant and $\mathcal{H}$ the CFF associated with the GPD $H$.

This leading power picture can be enhanced including kinematic power corrections that modify the hard kernel without involving new non-perturbative distributions \cite{Braun:2012hq,Braun:2012bg,Braun:2014sta,Braun:2016qlg,Braun:2020zjm,Braun:2022qly,Braun:2025xlp,Martinez-Fernandez:2025gub}.
These corrections are expected to be power suppressed in $1/\scale = 1/\sqrt{(Q^2 + t)}$, but explicitly depend on the spin of the target.
Their expression can be found in \cite{Braun:2014sta,Martinez-Fernandez:2025rcg}.

\section{Scalar targets and momentum distribution}

Following \cite{Martinez-Fernandez:2025rcg}, when taking into account kinematic power corrections, the DVCS subtraction constant is given for a (pseudo-)scalar target by:
\begin{align}
  \label{eq:h0++_tw4_full_scalar}
  \mathcal{S}^q(t,\scale^2) = & \int_{-1}^1d\alpha\  T_2^{++}(\alpha,t/\scale^2) D^q(\alpha,t) \nonumber\\
                & - 4\frac{M^2-t/4}{\scale^2}\int_\Omega \textrm{d}\Omega F^q(\beta,\alpha,t) \beta\,T_1^{++\,(1)}(\alpha)\,,
\end{align}
where $T_2^{++}$ and $T_1^{++\,(1)}$ are parts of the coefficient function. Their expressions are provided in the supplemental material.
Focusing on the second line, where the mass correction stands, we Taylor expand $T_1^{++\,(1)}(\alpha)$ around $\alpha = 0$:
\begin{align}
  \label{eq:Taylorexpansion}
  T_1^{++\,(1)}(\alpha) = \sum_{\substack{n=0\\\textrm{even}}}^\infty \alpha^n \mathfrak{c}_n = \sum_{\substack{n=0\\\textrm{even}}}^\infty \frac{\alpha^n}{n!} \left. \frac{\partial^n T_1^{++\,(1)}}{\partial \alpha^n}\right|_{\alpha = 0} .
\end{align}
Note that $T_1^{++\,(1)}$ is singular for $\alpha \to 1$. However, because of the support properties of DDs this singularity is tamed in Eq.~\eqref{eq:h0++_tw4_full_scalar} and the integrand goes to zero at the end-point.
Injecting Eq.~\eqref{eq:Taylorexpansion} into the second line of Eq.~\eqref{eq:h0++_tw4_full_scalar} and using Eqs.~\eqref{eq:DDF} and \eqref{eq:polynomialityH}, one gets:
\begin{align}
  \label{eq:MassTermSeries}
  \int_\Omega \textrm{d}\Omega F^q(\beta,\alpha) \beta\,T_1^{++\,(1)}(\alpha) = \mathfrak{c}_0 A_{1,0}^q + \sum_{\substack{n=2\\\textrm{even}}}^\infty \frac{2^n \mathfrak{c}_n}{n+1} A_{n+1,n}^q,
\end{align}
where the factor $2^n$ comes from the expansion in $(2\xi)^j$ in Eq. \eqref{eq:polynomialityH}.
The mass term can thus be expressed as a series of the subset $A_{n+1,n}$ of the generalised form factors $A_{i,j}$, whose first term $A^q_{1,0}$ is the EMT form factor containing the information regarding the momentum distribution within the nucleon.

We then truncate the series to the two first elements ($n\le 2$).
The expression of the subtraction constant becomes:
\begin{equation}
  \label{eq:ScalarDRTruncated}
   \mathcal{S}^q(t,\scale^2)  \hspace{-1mm} = \mathcal{D}^q(t,\scale^2) - 4\frac{M^2-\frac{t}{4}}{\scale^2}\hspace{-1mm}\left[\mathfrak{c}_0 A_{1,0}^q(t)+\hspace{-0.8mm}\frac{4 \mathfrak{c}_2 }{3}A_{3,2}^q(t) \hspace{-0.8mm}\right]\hspace{-1mm},
\end{equation}
with $\mathfrak{c}_0 \approx 0.864$ and $\mathfrak{c}_2 \approx 0.980$ computed from their definition in Eq. \eqref{eq:Taylorexpansion}, and
\begin{equation}
  \label{eq:IntD}
  \mathcal{D}^q(t,\scale^2) =  \int_{-1}^1 \textrm{d}\alpha\,T_2^{++}\left(\alpha,\frac{t}{\scale^2}\right) D^q(\alpha,t).
\end{equation}
In order to test the validity of this approximation, we introduce the ratio:
\begin{align}
  \label{eq:Rscalar}
  \mathcal{R}_n^q = \frac{\sum_{\substack{j=0\\\textrm{even}}}^n \frac{2^j \mathfrak{c}_j}{j+1} A_{j+1,j}^q}{\sum_{\substack{j=0\\\textrm{even}}}^\infty \frac{2^j \mathfrak{c}_j}{j+1} A_{j+1,j}^q},
\end{align}
and compute the $A_{n+1,n}$ generalised form factor from the so-called phenomenological model of Ref.~\cite{Chavez:2021llq}, which is based on the xFitter pion PDF \cite{Novikov:2020snp}, and the Radyushking DD Ansatz \cite{Musatov:1999xp}. 
\begin{table}[t]
  \centering
  \begin{tabular}{c||cc}
        $t\,[\mathrm{GeV}^2]$ & $\mathcal{R}_0(t)$ & $\mathcal{R}_2(t)$ \\ \hline\hline
        -0.0001  &   0.907   &  0.975 \\
        -0.3  &   0.923  &   0.982 \\
        -0.7  &   0.937  & 0.987 \\
        -1.2  &   0.949  &   0.991 \\
  \end{tabular}
  \caption{Values of $\mathcal{R} = \sum \eta_q \mathcal{R}^q$ with $\eta_q = e_q^2/(\sum e_{q'}^2)$ using the phenomenological pion model defined in Ref.~\cite{Chavez:2021llq}.}
  \label{tab:Rscalar}
\end{table}
The results, reported in Table~\ref{tab:Rscalar}, shows that the first term contributes to 90\% of the total sum, making the low-$n$ truncation an excellent approximation and thus justifying Eq. \eqref{eq:ScalarDRTruncated}. 
For (pseudo-)scalar targets, the DVCS subtraction constant can thus be seen as a relation between the EMT form factor $A$ and the Polyakov-Weiss $D$-term.

\section{DVCS dispersion relation and angular Momentum}

For spin-$\frac{1}{2}$ targets such as the nucleon, the situation is more complicated as the subtraction constant depends on both DDs $F$ and $K$ (see \cite{Braun:2014sta,Martinez-Fernandez:2025rcg}):
\begin{align}
  \label{eq:h0++_tw4_full_spin1/2}
  \mathcal{S}^q(t,\scale^2) = &\ \mathcal{D}^q(t,\scale^2)  - 4\frac{M^2}{\scale^2}\int_\Omega\, \textrm{d}\Omega \beta T_1^{++\,(1)}(\alpha)  \nonumber\\
                & \times \left[F^q(\beta,\alpha,t) + \frac{t}{4M^2}K^q(\beta,\alpha,t) \right] .
\end{align}
We proceed as previously, expanding $T_1^{++\,(1)}$ following Eq.~\eqref{eq:Taylorexpansion}.
Integrating over $F$ yields again a series of $A_{n+1,n}$ generalised form factors, see Eq.~\eqref{eq:MassTermSeries}.
Similarly, using Eqs. \eqref{eq:DDK} and \eqref{eq:polynomialityE}, the integral on $K$ can be expressed as:
\begin{align}
  \label{eq:KtermSeries}
  \int_\Omega \textrm{d}\Omega K^q(\beta,\alpha) \beta T_1^{++\,(1)}(\alpha) = \mathfrak{c}_0 B_{1,0}^q + \sum_{\substack{n=2\\\textrm{even}}}^\infty \frac{2^n \mathfrak{c}_n}{n+1} B^q_{n+1,n},
\end{align}
and we define the ratio $\mathcal{Y}_n$:
\begin{align}
  \label{eq:Yratio}
  \mathcal{Y}_n^q = \frac{\sum_{\substack{j=0\\\textrm{even}}}^n \frac{2^j \mathfrak{c}_j}{j+1} B_{j+1,j}^q}{\sum_{\substack{j=0\\\textrm{even}}}^\infty \frac{2^j \mathfrak{c}_j}{j+1} B_{j+1,j}^q},
\end{align}
assessing the precision of truncation of the $B$ series at order $n$. 

\begin{table}
  \centering
  \begin{tabular}{c||cccc}
    
    $t\,[\mathrm{GeV}^2]$ & $\mathcal{R}_0(t)$ & $\mathcal{R}_2(t)$ & $\mathcal{Y}_0(t)$ & $\mathcal{Y}_2(t)$ \\ \hline\hline
    -0.0001  &   0.844   &  0.947  &   0.815   &  0.930 \\
    -0.3  &   0.8556 &    0.953  &   0.833  &   0.941 \\
    -0.7   &  0.870   &  0.961   &  0.850   &  0.952 \\
    -1.2  &   0.885  &   0.969  &   0.866  &   0.960 \\
  \end{tabular}
  \caption{Values of the charge-weighted ratios $\mathcal{R}$ and $\mathcal{Y}$ using the Goloskokov-Kroll models of Refs.~\cite{Goloskokov:2006hr,Goloskokov:2007nt,Goloskokov:2008ib}.}
  \label{tab:intWithFandK}
\end{table}

Using the Goloskokov-Kroll model \cite{Goloskokov:2006hr,Goloskokov:2007nt,Goloskokov:2008ib}, we can assess, as done before, the validity of the low-$n$ truncation.
Note that the lattice computations remain exploratory and limited to the $u-d$ combination \cite{HadStruc:2024rix}.
The results, shown in Tab.~\ref{tab:intWithFandK}, are similar to the scalar case, with the first term of the $A$ and $B$ series contributing to 82\% of the total contribution. 

Now recalling that the total angular momentum of the nucleon $J^q$ is given as:
\begin{align}
  \label{eq:Jdef}
  J^q(t) = \frac{A_{1,0}^q(t) + B_{1,0}^q (t)}{2},
\end{align}
one can express the nucleon DVCS subtraction constant as:
\begin{align}
  \label{eq:NucleonFull}
  \mathcal{S}^q(t,\scale^2) = &\ \mathcal{D}^q(t,\scale^2) - 4\frac{M^2}{\scale^2}\mathfrak{c}_0\left[A^q_{1,0}(t) + \frac{t}{4M^2}B_{1,0}^q(t) \right] \nonumber \\
                 &  - 4\frac{M^2}{\scale^2}\hspace{-1mm} \sum_{\substack{n=2\\\textrm{even}}}^\infty \hspace{-1mm}\frac{2^n \mathfrak{c}_n}{n+1}\hspace{-1mm}\left[A_{n+1,n}^q(t)+\frac{tB_{n+1,n}^q(t)}{4M^2} \right] \nonumber \\
  \approx &\ \mathcal{D}^q(t,\scale^2)  \hspace{-1mm}-\hspace{-1mm} \frac{4M^2\mathfrak{c}_0}{\scale^2}\hspace{-1mm}\left[\hspace{-0.8 mm}\frac{4M^2-t}{4M^2}A^q_{1,0}(t) \hspace{-1mm}+\hspace{-1mm} \frac{t}{2M^2}J^q(t)\hspace{-0.8mm} \right]\hspace{-1mm}.
\end{align}
Thus, in the nucleon case, the DVCS subtraction constant provides a relation between the Polyakov-Weiss $D$-term, the momentum distribution and the total angular momentum one.
Assuming that the first coefficient of the $D$-term expansion on Gegenbauer polynomials basis is also the dominant one (this is a standard assumption regularising the deconvolution problem, although its validity remains questionable \cite{Dutrieux:2021nlz,Dutrieux:2024bgc}), one can obtain the approximated relation:
\begin{align}
  \label{eq:GFFapprox}
  \mathcal{S}^q(t,\scale^2) \approx &\ 20 C^q(t)\left(1-\frac{t}{3\scale^2}\right) \nonumber \\
  & - \frac{4M^2\mathfrak{c}_0}{\scale^2}\left[\frac{4M^2-t}{4M^2}A^q_{1,0}(t) + \frac{t}{2M^2}J^q(t)\right],
\end{align}
connecting the EMT form factors. 
This holds in Fourier space: the maps of pressure and shear forces within the nucleon depend on the maps of momentum and angular momentum.

\section{Assessing the DVCS Dispersion relation}

We are now in a position to assess the size of the kinematics power corrections at achievable DVCS kinematics.
To do that, we will exploit the lattice computations of the EMT form factors of Ref.~\cite{Hackett:2023rif} (using the dipole fit), and the Continuum Schwinger Method (CSM) computation of Ref.~\cite{Yao:2024ixu}.
We then compute the total subtraction constant:
\begin{align}
  \label{eq:TotalS}
  \mathcal{S} = \sum_q e_q^2 \mathcal{S}^q .
\end{align}
The lattice results\footnote{Note the factor 4 difference in the definition of $C$ between our convention and refs. \cite{Hackett:2023rif,Yao:2024ixu}.} are provided at a renormalisation scale of $\mu^2_{\textrm{ref.}} = 4\textrm{GeV}^2$, and we evolve them to $\mu^2 = 2 \mathrm{GeV^2}$ using Leading-Order (LO) evolution equations.
The choice of this scale is driven by the experimental knowledge of the subtraction constant, which is maximal for $Q^2 \approx 2 \textrm{GeV}^2$ and $t \approx -0.4\textrm{GeV}^2$ (see Ref.~\cite{Dutrieux:2024bgc}).
Note that this is a bit below the threshold of $\scale_{\textrm{th}}^2 = 2M^2$ above which the dispersion formalism is proven to work \cite{Dutrieux:2024bgc,Martinez-Fernandez:2025rcg}, but (i) no breakdown of dispersion relation is seen in experimental data, and (ii) a fine tuning of the values of $t$ and $Q^2$ would not change our discussion.  
\begin{figure}[t]
  \centering
  \includegraphics[width=0.48\textwidth]{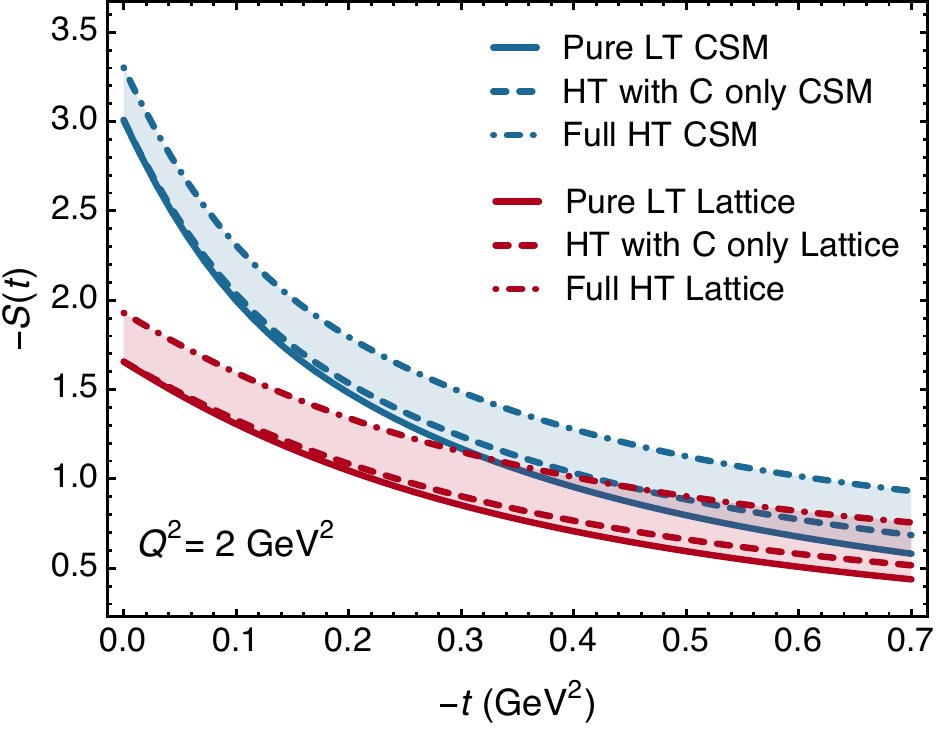}\\
  \includegraphics[width=0.48\textwidth]{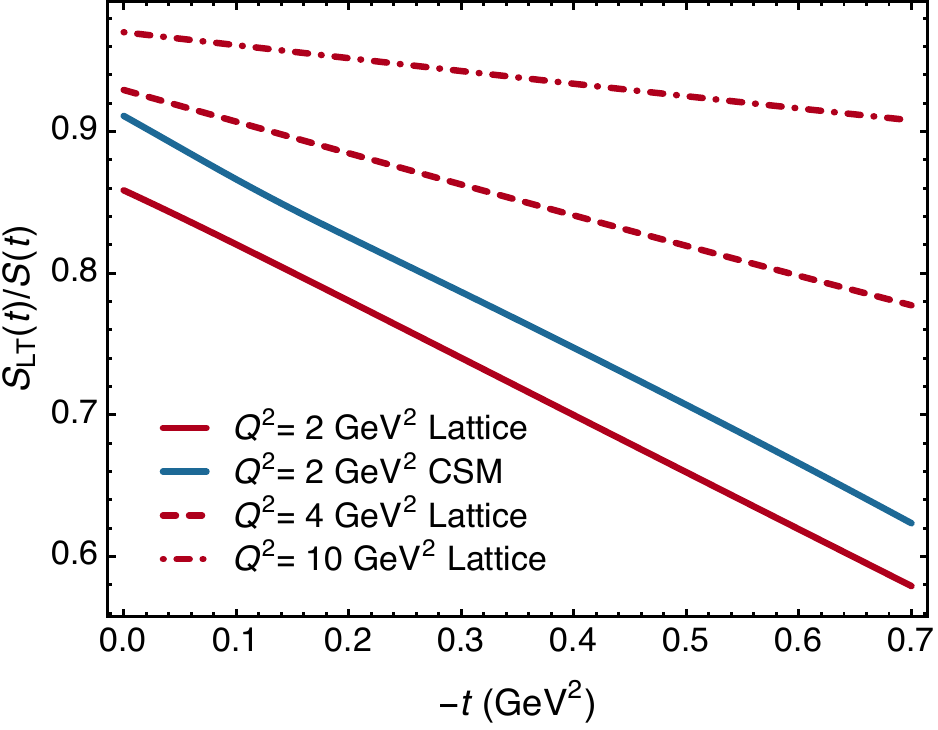}
  \caption{
Upper panel: evaluation of the DVCS subtraction constant for $Q^2 = \mu^2 = 2\mathrm{GeV^2}$ using three quark flavours. CSM  (lattice) calculations are shown by blue (red) curves. Continuous lines represent results obtained within the leading twist formalism; dashed lines include power corrections neglecting the second line of Eq.~\eqref{eq:GFFapprox}; finally, dashed-dotted lines include the full twist-4 correction using Eq.~\eqref{eq:GFFapprox}. The impact of HT corrections is highlighted by the corresponding band in both cases.
    Lower Panel: ratio between the LT description of the subtraction constant and Eq.~\eqref{eq:GFFapprox}.}
  \label{fig:Sq2GeV2}
\end{figure}

The results of the computation of the DVCS subtraction constant are shown on Fig.~\ref{fig:Sq2GeV2}.
For both approaches, the $M^2/\scale^2$ part of the corrections largely dominates the $t/\scale^2$ ones. The total power correction account for $1/3$ (lattice) and $1/4$ (CSM) effect at $Q^2 = 2\textrm{GeV}^2$ and $t=-0.4\textrm{GeV}^2$.
In addition, $J^q$ and $C^q$ have opposite signs, triggering a compensation between $t C^q /\scale^2$ and $t J^q /\scale^2$, and explaining also why the $t/\scale^2$ effect is small.
The difference at small-$t$ between the two approaches comes from the fact that, despite having compatible total $C$ form factor, the decomposition between quarks and gluons is significantly different with $C^g_{\textrm{Latt}} \approx 4.5 C^u_{\textrm{Latt}}$ while $C^g_{\textrm{CSM}} \approx C_{\textrm{CSM}}^u$.
At $Q^2 = 4\mathrm{GeV^2}$, the impact of the power corrections is reduced to roughly 15\%. This is due to the power-law suppression, but also to the decrease of $A^q$, as more momentum is sent to the gluon sector.

\section{Conclusion}

The kinematic power corrections contribute significantly to the DVCS dispersion relation.
They thus modify our understanding of the subtraction constant of DVCS, which now connects the EMT form factors $A^q$ and $J^q$ with $C^q$, although the contribution of $-t J^q/\scale^2$ is assessed here to be much smaller than the one of $M^2 A^q/\scale^2 $, with opposite sign compared to $-tC^q/\scale^2$ and $tA^q/\scale^2$. Since according to lattice computations $J^q(0)< A^q(0) < -C^q(0)$, an experimental extraction through a deviation of the $t$-slope will be challenging.

At $Q^2 = 2\textrm{\ GeV}^2$, i.e.~the highest $Q^2$ range of the JLab 6 GeV kinematical area, the contribution of $A$ \emph{is} sizeable, as it can account for $1/3$ of the experimental signal, and largely dominates the power correction effects. 
A 30\% to 40\% effect is in agreement with the expectations that have been made previously on DVCS amplitudes~\cite{Defurne:2015kxq,Braun:2025xlp}. 

This outweighs the NLO corrections obtained previously \cite{Diehl:2007jb,Dutrieux:2024bgc} (with the caveat of the large gluon component computed on the lattice), but does not resolve the deconvolution problem of the $D$-term \cite{Martinez-Fernandez:2025rcg}.

However, it confirms what some of us already suspected, that interpreting correctly JLab 6GeV data is really challenging, as the low-$Q^2$ values do not guarantee that power corrections are actually suppressed.
In that respect, higher $Q^2$ data at JLab 12 and JLab 20 are more than welcome.
Note that since a significant part of EIC data will be taken at $Q^2$ of the order of a few $\textrm{GeV}^2$, kinematic power corrections remain a relevant topic there too. 

Finally let us point out that the uncertainty on the subtraction constant remains large, mostly because the real parts of the CFF are poorly known.
A positron beam at JLab or the EIC would allow one to better constrain the latter, improving significantly the extraction of the DVCS subtraction constant. 

\begin{acknowledgments}
  The authors would like to thank V. Bertone, H. Dutrieux, C. Lorcé, B. Pire and P. Sznajder for stimulating discussion.
  This research was funded in part by l’Agence Nationale de la Recherche (ANR), project ANR-23-CE31-0019.
  For the purpose of open access, the author has applied a CC-BY public copyright licence to any Author Accepted Manuscript (AAM) version arising from this submission.
  This work was made possible by Institut Pascal at Université Paris-Saclay with the support of the program “Investissements d’avenir” ANR-11-IDEX-0003-01.
\end{acknowledgments}

\appendix
\begin{widetext}
  \section{Hard coefficient functions of the subtraction constant associated to quark GPDs $H$ and $E$}
  
  In accordance to Ref.~\cite{Martinez-Fernandez:2025rcg}, both spin-0 and spin-1/2 targets make use of the same three functions $T_0^{++}$ and $T_1^{++}$ for building up the CFF $\cffH^{++}$ of GPD $H$ and its associated subtraction constant for the dispersion relation. For a spin-1/2, this is also true for $\cffE^{++}$ of GPD $E$. Note, however, that the form the subtraction is spin-dependent as explained in Ref.~\cite{Martinez-Fernandez:2025rcg}. These hard coefficient functions are thus
  \begin{align}
    \label{eq:T0++_and_T1++}
    T_0^{++}\left(\frac{x}{\xi}, \frac{t}{\scale^2}\right) = &\ C_{\rm LT}^{(+)}\left(\frac{x}{\xi}\right) +  \frac{t}{\scale^2}\wpbbiii^{(+)}\left(\frac{x}{\xi}\right) - \frac{t}{\scale^2}\frac{\mathcal{L}^{(+)}\left(\frac{x}{\xi}\right)+C_0^{(+)}\left(\frac{x}{\xi}\right)}{2} \,,\\
    T_1^{++}\left(\frac{x}{\xi}\right)  = &\frac{\mathcal{L}^{(+)}\left(\frac{x}{\xi}\right) - \wpbbiii^{(+)}\left(\frac{x}{\xi}\right)}{2}\,.
  \end{align}
  Here, all the kernels in the RHS are antisymmetric. In particular, $C_{\rm LT}^{(+)}$ is the only one surviving the leading-twist limit, can be found at next-to-leading order (NLO) in $\alpha_s$ in Ref.~\cite{Diehl:2007jb}, and $C_0^{(+)}$ is its LO component. The other two functions come from a twist calculation~\cite{Martinez-Fernandez:2025gub} and read
  \begin{align}
    \wpbbiii^{(+)}(x/\xi) & = \frac{-2}{x/\xi+1}\Ln{\frac{x/\xi-1+i0}{-2+i0}} - (x/\xi\to -x/\xi) \,, \label{eq:wpbbiii} \\
    \mathcal{L}^{(+)}(x/\xi) & = \frac{4}{x/\xi-1}\left[ \Li{2}{\frac{x/\xi+1}{2-i0}} - \Li{2}{1} \right] - (x/\xi\to -x/\xi)\,. \label{eq:calL}
  \end{align}
  In the part of the subtraction constant corresponding to the $D$-term, the kernel ($T_2^{++}$, in notation of the main text) is a combination of the above two:
  \begin{equation}\label{eq:T_2++}
    T_2^{++}(\alpha, t/\scale^2) = T_0^{++}(\alpha,t/\scale^2) + \frac{t}{\scale^2}T_1^{++}(\alpha) \,.
  \end{equation}
  The novel term including the DD $F$ for a spin-0 target, and both $F$ and $K$ for the spin-1/2 case, is the first derivative of $T_1^{++}$, this is
  \begin{align}
    T_1^{++\,(1)}(\alpha) \overset{\alpha\in(-1,1)}{=} &\ \Re\ 
                                                         \left( T_1^{++\,(1)}(\alpha) \right) \nonumber\\
    \overset{\alpha\in(-1,1)}{=} &\ \frac{1+3\alpha}{(1+\alpha)^2(1-\alpha)}\Ln{\frac{1-\alpha}{2}} - \frac{2}{(1-\alpha)^2}\left[ \Li{2}{\frac{1+\alpha}{2}} - \Li{2}{1} \right] - \frac{1}{1-\alpha^2} \nonumber\\
                                                       &\ + (\alpha\to -\alpha) \,,
  \end{align}
  where $\Re$ stands for the real part of the function.
\end{widetext}


\bibliographystyle{unsrt}
\bibliography{Bibliography}

@Article{Yao:2024ixu,
  author        = {Yao, Z. -Q. and Xu, Y. -Z. and Binosi, D. and Cui, Z. -F. and Ding, M. and Raya, K. and Roberts, C. D. and Rodr{\'\i}guez-Quintero, J. and Schmidt, S. M.},
  title         = {{Nucleon gravitational form factors}},
  journal       = {Eur. Phys. J. A},
  year          = {2025},
  volume        = {61},
  number        = {5},
  pages         = {92},
  archiveprefix = {arXiv},
  doi           = {10.1140/epja/s10050-025-01557-x},
  eprint        = {2409.15547},
  primaryclass  = {hep-ph},
  reportnumber  = {NJU-INP 091/24},
}

@Article{Radyushkin:1998bz,
  Title                    = {{Symmetries and structure of skewed and double distributions}},
  Author                   = {Radyushkin, A.V.},
  Journal                  = {Phys.Lett.},
  Year                     = {1999},
  Pages                    = {81-88},
  Volume                   = {B449},

  Archiveprefix            = {arXiv},
  Doi                      = {10.1016/S0370-2693(98)01584-6},
  Eprint                   = {hep-ph/9810466},
  Primaryclass             = {hep-ph},
  Reportnumber             = {JLAB-THY-98-41},
  Slaccitation             = {%%CITATION = HEP-PH/9810466;%%}
}

@Article{Radyushkin:1997ki,
  Title                    = {{Nonforward parton distributions}},
  Author                   = {Radyushkin, A.V.},
  Journal                  = {Phys.Rev.},
  Year                     = {1997},
  Pages                    = {5524-5557},
  Volume                   = {D56},

  Archiveprefix            = {arXiv},
  Doi                      = {10.1103/PhysRevD.56.5524},
  Eprint                   = {hep-ph/9704207},
  Primaryclass             = {hep-ph},
  Reportnumber             = {JLAB-THY-97-10},
  Slaccitation             = {%%CITATION = HEP-PH/9704207;%%}
}

@Article{Polyakov:1999gs,
  Title                    = {{Skewed and double distributions in pion and nucleon}},
  Author                   = {Polyakov, Maxim V. and Weiss, C.},
  Journal                  = {Phys.Rev.},
  Year                     = {1999},
  Pages                    = {114017},
  Volume                   = {D60},

  Archiveprefix            = {arXiv},
  Doi                      = {10.1103/PhysRevD.60.114017},
  Eprint                   = {hep-ph/9902451},
  Primaryclass             = {hep-ph},
  Reportnumber             = {RUB-TPII-1-99},
  Slaccitation             = {%%CITATION = HEP-PH/9902451;%%}
}

@Article{Polyakov:2002yz,
  Title                    = {{Generalized parton distributions and strong forces inside nucleons and nuclei}},
  Author                   = {Polyakov, M. V.},
  Journal                  = {Phys. Lett.},
  Year                     = {2003},
  Pages                    = {57-62},
  Volume                   = {B555},

  Archiveprefix            = {arXiv},
  Doi                      = {10.1016/S0370-2693(03)00036-4},
  Eprint                   = {hep-ph/0210165},
  Primaryclass             = {hep-ph},
  Reportnumber             = {RUB-TP2-14-02},
  Slaccitation             = {%%CITATION = HEP-PH/0210165;%%}
}

@Article{Novikov:2020snp,
  author        = {Novikov, I. and others},
  journal       = {Phys. Rev. D},
  title         = {{Parton Distribution Functions of the Charged Pion Within The xFitter Framework}},
  year          = {2020},
  number        = {1},
  pages         = {014040},
  volume        = {102},
  archiveprefix = {arXiv},
  doi           = {10.1103/PhysRevD.102.014040},
  eprint        = {2002.02902},
  primaryclass  = {hep-ph},
  reportnumber  = {DESY-20-013, DESY 20-013},
}

@Article{Musatov:1999xp,
  Title                    = {{Evolution and models for skewed parton distributions}},
  Author                   = {Musatov, I.V. and Radyushkin, A.V.},
  Journal                  = {Phys.Rev.},
  Year                     = {2000},
  Pages                    = {074027},
  Volume                   = {D61},

  Archiveprefix            = {arXiv},
  Doi                      = {10.1103/PhysRevD.61.074027},
  Eprint                   = {hep-ph/9905376},
  Primaryclass             = {hep-ph},
  Reportnumber             = {JLAB-THY-99-12},
  Slaccitation             = {%%CITATION = HEP-PH/9905376;%%}
}

@Article{Mueller:1998fv,
  Title                    = {{Wave functions, evolution equations and evolution kernels from light ray operators of QCD}},
  Author                   = {Mueller, Dieter and Robaschik, D. and Geyer, B. and Dittes, F.M. and Ho\v{r}e\v{j}si, J.},
  Journal                  = {Fortsch.Phys.},
  Year                     = {1994},
  Pages                    = {101-141},
  Volume                   = {42},

  Archiveprefix            = {arXiv},
  Doi                      = {10.1002/prop.2190420202},
  Eprint                   = {hep-ph/9812448},
  Primaryclass             = {hep-ph},
  Reportnumber             = {NTZ-6-91},
  Slaccitation             = {%%CITATION = HEP-PH/9812448;%%}
}

@Article{Mezrag:2022pqk,
  author        = {Mezrag, C\'edric},
  journal       = {Few Body Syst.},
  title         = {{An Introductory Lecture on Generalised Parton Distributions}},
  year          = {2022},
  number        = {3},
  pages         = {62},
  volume        = {63},
  archiveprefix = {arXiv},
  doi           = {10.1007/s00601-022-01765-x},
  eprint        = {2207.13584},
  primaryclass  = {hep-ph},
}

@Article{Martinez-Fernandez:2025gub,
  author        = {Martinez-Fernandez, V. and Pire, B. and Sznajder, P. and Wagner, J.},
  title         = {{Timelike Compton scattering on a spin-0 target with kinematic twist-4 precision}},
  journal       = {Phys. Rev. D},
  year          = {2025},
  volume        = {111},
  number        = {7},
  pages         = {074034},
  archiveprefix = {arXiv},
  doi           = {10.1103/PhysRevD.111.074034},
  eprint        = {2503.02461},
  primaryclass  = {hep-ph},
  reportnumber  = {CPHT-RR088.122024},
}

@Article{Martinez-Fernandez:2025rcg,
  author        = {Mart{\'\i}nez-Fern{\'a}ndez, V{\'\i}ctor and Mezrag, C{\'e}dric},
  title         = {{Dispersion relations of deeply virtual Compton scattering: investigating twist-4 kinematic power corrections}},
  journal       = {arXiv:2509.05059},
  year          = {2025},
  archiveprefix = {arXiv},
  eprint        = {2509.05059},
  primaryclass  = {hep-ph},
}

@Article{Lorce:2025oot,
  author        = {Lorc{\'e}, C{\'e}dric and Schweitzer, Peter},
  title         = {{Pressure inside hadrons: criticism, conjectures, and all that}},
  journal       = {Acta Phys. Polon. B},
  year          = {2025},
  volume        = {56},
  pages         = {3--A17},
  archiveprefix = {arXiv},
  doi           = {10.5506/APhysPolB.56.3-A17},
  eprint        = {2501.04622},
  primaryclass  = {hep-ph},
}

@Article{Kumericki:2019ddg,
  author       = {Kumerički, Krešimir},
  title        = {{Measurability of pressure inside the proton}},
  journal      = {Nature},
  year         = {2019},
  volume       = {570},
  number       = {7759},
  pages        = {E1-E2},
  doi          = {10.1038/s41586-019-1211-6},
  slaccitation = {%%CITATION = NATUA,570,E1;%%},
}

@Article{Ji:1998pc,
  Title                    = {{Off forward parton distributions}},
  Author                   = {Ji, Xiang-Dong},
  Journal                  = {J.Phys.},
  Year                     = {1998},
  Pages                    = {1181-1205},
  Volume                   = {G24},

  Archiveprefix            = {arXiv},
  Doi                      = {10.1088/0954-3899/24/7/002},
  Eprint                   = {hep-ph/9807358},
  Primaryclass             = {hep-ph},
  Reportnumber             = {UMD-PP-98-092, DOE-ER-40762-144},
  Slaccitation             = {%%CITATION = HEP-PH/9807358;%%}
}

@Article{Ji:1996ek,
  Title                    = {{Gauge-Invariant Decomposition of Nucleon Spin}},
  Author                   = {Ji, Xiang-Dong},
  Journal                  = {Phys. Rev. Lett.},
  Year                     = {1997},
  Pages                    = {610-613},
  Volume                   = {78},

  Archiveprefix            = {arXiv},
  Doi                      = {10.1103/PhysRevLett.78.610},
  Eprint                   = {hep-ph/9603249},
  Primaryclass             = {hep-ph},
  Reportnumber             = {MIT-CTP-2517},
  Slaccitation             = {%%CITATION = HEP-PH/9603249;%%}
}

@Article{Ji:1996nm,
  Title                    = {{Deeply virtual Compton scattering}},
  Author                   = {Ji, Xiang-Dong},
  Journal                  = {Phys.Rev.},
  Year                     = {1997},
  Pages                    = {7114-7125},
  Volume                   = {D55},

  Archiveprefix            = {arXiv},
  Doi                      = {10.1103/PhysRevD.55.7114},
  Eprint                   = {hep-ph/9609381},
  Primaryclass             = {hep-ph},
  Reportnumber             = {UMD-PP-97-26, MIT-CTP-2568},
  Slaccitation             = {%%CITATION = HEP-PH/9609381;%%}
}

@Article{Ji:2025qax,
  author        = {Ji, Xiangdong and Yang, Chen},
  title         = {{A Journey of Seeking Pressures and Forces in the Nucleon}},
  journal       = {arxiv:2508.16727},
  year          = {2025},
  archiveprefix = {arXiv},
  eprint        = {2508.16727},
  primaryclass  = {hep-ph},
}

@Article{Hackett:2023rif,
  author        = {Hackett, Daniel C. and Pefkou, Dimitra A. and Shanahan, Phiala E.},
  title         = {{Gravitational Form Factors of the Proton from Lattice QCD}},
  journal       = {Phys. Rev. Lett.},
  year          = {2024},
  volume        = {132},
  number        = {25},
  pages         = {251904},
  archiveprefix = {arXiv},
  doi           = {10.1103/PhysRevLett.132.251904},
  eprint        = {2310.08484},
  primaryclass  = {hep-lat},
  reportnumber  = {MIT-CTP/5630, FERMILAB-PUB-23-592-T},
}

@Article{Goloskokov:2008ib,
  author        = {Goloskokov, S. V. and Kroll, P.},
  title         = {{The Target asymmetry in hard vector-meson electroproduction and parton angular momenta}},
  journal       = {Eur. Phys. J. C},
  year          = {2009},
  volume        = {59},
  pages         = {809--819},
  archiveprefix = {arXiv},
  doi           = {10.1140/epjc/s10052-008-0833-x},
  eprint        = {0809.4126},
  primaryclass  = {hep-ph},
  reportnumber  = {WU-B-08-05, WU B 08-05},
}

@Article{Goloskokov:2007nt,
  Title                    = {{The Role of the quark and gluon GPDs in hard vector-meson electroproduction}},
  Author                   = {Goloskokov, S.V. and Kroll, P.},
  Journal                  = {Eur.Phys.J.},
  Year                     = {2008},
  Pages                    = {367-384},
  Volume                   = {C53},

  Archiveprefix            = {arXiv},
  Doi                      = {10.1140/epjc/s10052-007-0466-5},
  Eprint                   = {0708.3569},
  Primaryclass             = {hep-ph},
  Reportnumber             = {WU-B-07-07},
  Slaccitation             = {%%CITATION = ARXIV:0708.3569;%%}
}

@Article{Goloskokov:2006hr,
  Title                    = {{The Longitudinal cross-section of vector meson electroproduction}},
  Author                   = {Goloskokov, S.V. and Kroll, P.},
  Journal                  = {Eur.Phys.J.},
  Year                     = {2007},
  Pages                    = {829-842},
  Volume                   = {C50},

  Archiveprefix            = {arXiv},
  Doi                      = {10.1140/epjc/s10052-007-0228-4},
  Eprint                   = {hep-ph/0611290},
  Primaryclass             = {hep-ph},
  Reportnumber             = {WU-B-06-02},
  Slaccitation             = {%%CITATION = HEP-PH/0611290;%%}
}

@Article{Dutrieux:2024bgc,
  author        = {Dutrieux, Herv{\'e} and Meisgny, Thibaud and Mezrag, C{\'e}dric and Moutarde, Herv{\'e}},
  title         = {{Proton internal pressure from deeply virtual Compton scattering on collider kinematics}},
  journal       = {Eur. Phys. J. C},
  year          = {2025},
  volume        = {85},
  number        = {1},
  pages         = {105},
  archiveprefix = {arXiv},
  doi           = {10.1140/epjc/s10052-024-13737-y},
  eprint        = {2410.13518},
  primaryclass  = {hep-ph},
}

@Article{Dutrieux:2021nlz,
  author        = {Dutrieux, H. and Lorc\'e, C. and Moutarde, H. and Sznajder, P. and Trawi\'nski, A. and Wagner, J.},
  journal       = {Eur. Phys. J. C},
  title         = {{Phenomenological assessment of proton mechanical properties from deeply virtual Compton scattering}},
  year          = {2021},
  number        = {4},
  pages         = {300},
  volume        = {81},
  archiveprefix = {arXiv},
  doi           = {10.1140/epjc/s10052-021-09069-w},
  eprint        = {2101.03855},
  primaryclass  = {hep-ph},
}

@article{HadStruc:2024rix,
    author = "Dutrieux, Herv\'e and Edwards, Robert G. and Egerer, Colin and Karpie, Joseph and Monahan, Christopher and Orginos, Kostas and Radyushkin, Anatoly and Richards, David and Romero, Eloy and Zafeiropoulos, Savvas",
    collaboration = "HadStruc",
    title = "{Towards unpolarized GPDs from pseudo-distributions}",
    eprint = "2405.10304",
    archivePrefix = "arXiv",
    primaryClass = "hep-lat",
    reportNumber = "JLAB-THY-24-4059, JLAB-THY-24-4059",
    doi = "10.1007/JHEP08(2024)162",
    journal = "JHEP",
    volume = "08",
    pages = "162",
    year = "2024"
}

@Article{Diehl:2007jb,
  author        = {Diehl, M. and Ivanov, D. Yu.},
  journal       = {Eur. Phys. J. C},
  title         = {{Dispersion representations for hard exclusive processes: beyond the Born approximation}},
  year          = {2007},
  pages         = {919--932},
  volume        = {52},
  archiveprefix = {arXiv},
  doi           = {10.1140/epjc/s10052-007-0401-9},
  eprint        = {0707.0351},
  primaryclass  = {hep-ph},
  reportnumber  = {DESY-07-094},
}

@Article{Defurne:2015kxq,
  Title                    = {{E00-110 experiment at Jefferson Lab Hall A: Deeply virtual Compton scattering off the proton at 6 GeV}},
  Author                   = {Defurne, M. and others},
  Journal                  = {Phys. Rev.},
  Year                     = {2015},
  Number                   = {5},
  Pages                    = {055202},
  Volume                   = {C92},

  Archiveprefix            = {arXiv},
  Collaboration            = {Jefferson Lab Hall A},
  Doi                      = {10.1103/PhysRevC.92.055202},
  Eprint                   = {1504.05453},
  Primaryclass             = {nucl-ex},
  Reportnumber             = {IRFU-15-12, JLAB-PHY-15-2038},
  Slaccitation             = {%%CITATION = ARXIV:1504.05453;%%}
}

@Article{Chouika:2017dhe,
  Title                    = {{Covariant Extension of the GPD overlap representation at low Fock states}},
  Author                   = {Chouika, N. and Mezrag, C. and Moutarde, H. and Rodríguez-Quintero, J.},
  Journal                  = {Eur. Phys. J.},
  Year                     = {2017},
  Pages                    = {906},
  Volume                   = {C77},

  Archiveprefix            = {arXiv},
  Doi                      = {10.1140/epjc/s10052-017-5465-6},
  Eprint                   = {1711.05108},
  Primaryclass             = {hep-ph},
  Slaccitation             = {%%CITATION = ARXIV:1711.05108;%%}
}

@Article{Chavez:2021llq,
  author        = {Chavez, Jos\'e Manuel Morgado and Bertone, Valerio and De Soto Borrero, Feliciano and Defurne, Maxime and Mezrag, C\'edric and Moutarde, Herv\'e and Rodr\'\i{}guez-Quintero, Jos\'e and Segovia, Jorge},
  journal       = {Phys. Rev. D},
  title         = {{Pion generalized parton distributions: A path toward phenomenology}},
  year          = {2022},
  number        = {9},
  pages         = {094012},
  volume        = {105},
  archiveprefix = {arXiv},
  doi           = {10.1103/PhysRevD.105.094012},
  eprint        = {2110.06052},
  primaryclass  = {hep-ph},
}

@article{Burkert:2018bqq,
    author = "Burkert, V. D. and Elouadrhiri, L. and Girod, F. X.",
    title = "{The pressure distribution inside the proton}",
    doi = "10.1038/s41586-018-0060-z",
    year = "2018",
    journal = "Nature",
    volume = "557",
    number = "7705",
    pages = "396--399"
}

@Article{Braun:2014sta,
  author        = {Braun, Vladimir M. and Manashov, Alexander N. and M{\"u}ller, Dieter and Pirnay, Bjoern M.},
  title         = {{Deeply Virtual Compton Scattering to the twist-four accuracy: Impact of finite-$t$ and target mass corrections}},
  journal       = {Phys. Rev. D},
  year          = {2014},
  volume        = {89},
  number        = {7},
  pages         = {074022},
  archiveprefix = {arXiv},
  doi           = {10.1103/PhysRevD.89.074022},
  eprint        = {1401.7621},
  primaryclass  = {hep-ph},
}

@Article{Braun:2016qlg,
  author        = {Braun, V. M. and Manashov, A. N. and Moch, S. and Strohmaier, M.},
  title         = {{Two-loop conformal generators for leading-twist operators in QCD}},
  journal       = {JHEP},
  year          = {2016},
  volume        = {03},
  pages         = {142},
  archiveprefix = {arXiv},
  doi           = {10.1007/JHEP03(2016)142},
  eprint        = {1601.05937},
  primaryclass  = {hep-ph},
  reportnumber  = {DESY-16-012},
}

@Article{Braun:2020zjm,
  author        = {Braun, V. M. and Ji, Yao and Manashov, A. N.},
  title         = {{Two-photon processes in conformal QCD: resummation of the descendants of leading-twist operators}},
  journal       = {JHEP},
  year          = {2021},
  volume        = {03},
  pages         = {051},
  archiveprefix = {arXiv},
  doi           = {10.1007/JHEP03(2021)051},
  eprint        = {2011.04533},
  primaryclass  = {hep-ph},
  reportnumber  = {DESY 20-189, DESY-20-189, SI-HEP-2020-26},
}

@Article{Braun:2022qly,
  author        = {Braun, V. M. and Ji, Yao and Manashov, A. N.},
  title         = {{Next-to-leading-power kinematic corrections to DVCS: a scalar target}},
  journal       = {JHEP},
  year          = {2023},
  volume        = {01},
  pages         = {078},
  archiveprefix = {arXiv},
  doi           = {10.1007/JHEP01(2023)078},
  eprint        = {2211.04902},
  primaryclass  = {hep-ph},
  reportnumber  = {TUM-HEP-1432/22, DESY-22-169},
}

@Article{Braun:2025xlp,
  author        = {Braun, V. M. and Ji, Yao and Manashov, A. N.},
  title         = {{Kinematic power corrections to deeply virtual Compton scattering to twist-six accuracy}},
  journal       = {Phys. Rev. D},
  year          = {2025},
  volume        = {111},
  number        = {7},
  pages         = {076011},
  archiveprefix = {arXiv},
  doi           = {10.1103/PhysRevD.111.076011},
  eprint        = {2501.08185},
  primaryclass  = {hep-ph},
  reportnumber  = {TUM-HEP-1551/25, DESY 24--223},
}

@Article{Braun:2012bg,
  Title                    = {{Finite-t and target mass corrections to DVCS on a scalar target}},
  Author                   = {Braun, V.M. and Manashov, A.N. and Pirnay, B.},
  Journal                  = {Phys.Rev.},
  Year                     = {2012},
  Pages                    = {014003},
  Volume                   = {D86},

  Archiveprefix            = {arXiv},
  Doi                      = {10.1103/PhysRevD.86.014003},
  Eprint                   = {1205.3332},
  Primaryclass             = {hep-ph},
  Reportnumber             = {IPHT-T12-038},
  Slaccitation             = {%%CITATION = ARXIV:1205.3332;%%}
}

@Article{Braun:2012hq,
  Title                    = {{Finite-t and target mass corrections to deeply virtual Compton scattering}},
  Author                   = {Braun, V.M. and Manashov, A.N. and Pirnay, B.},
  Journal                  = {Phys.Rev.Lett.},
  Year                     = {2012},
  Pages                    = {242001},
  Volume                   = {109},

  Archiveprefix            = {arXiv},
  Doi                      = {10.1103/PhysRevLett.109.242001},
  Eprint                   = {1209.2559},
  Primaryclass             = {hep-ph},
  Slaccitation             = {%%CITATION = ARXIV:1209.2559;%%}
}

@Article{Belitsky:2012ch,
  Title                    = {{Compton scattering: from deeply virtual to quasi-real}},
  Author                   = {Belitsky, Andrei V. and Müller, Dieter and Ji, Yao},
  Journal                  = {Nucl.Phys.},
  Year                     = {2014},
  Pages                    = {214-268},
  Volume                   = {B878},

  Archiveprefix            = {arXiv},
  Doi                      = {10.1016/j.nuclphysb.2013.11.014},
  Eprint                   = {1212.6674},
  Primaryclass             = {hep-ph},
  Slaccitation             = {%%CITATION = ARXIV:1212.6674;%%}
}

@Article{Belitsky:2001ns,
  Title                    = {{Theory of deeply virtual Compton scattering on the nucleon}},
  Author                   = {Belitsky, Andrei V. and Mueller, Dieter and Kirchner, A.},
  Journal                  = {Nucl.Phys.},
  Year                     = {2002},
  Pages                    = {323-392},
  Volume                   = {B629},

  Archiveprefix            = {arXiv},
  Doi                      = {10.1016/S0550-3213(02)00144-X},
  Eprint                   = {hep-ph/0112108},
  Primaryclass             = {hep-ph},
  Reportnumber             = {DOE-ER-40762-009, UMD-PP-02-011, YITP-SB-01-51},
  Slaccitation             = {%%CITATION = HEP-PH/0112108;%%}
}

@Article{Anikin:2007yh,
  author        = {Anikin, I. V. and Teryaev, O. V.},
  journal       = {Phys. Rev. D},
  title         = {{Dispersion relations and subtractions in hard exclusive processes}},
  year          = {2007},
  pages         = {056007},
  volume        = {76},
  archiveprefix = {arXiv},
  doi           = {10.1103/PhysRevD.76.056007},
  eprint        = {0704.2185},
  primaryclass  = {hep-ph},
}

\end{document}